\documentclass[11pt]{article}
\usepackage{graphicx}
\usepackage{amsmath}
\usepackage{amssymb}
\usepackage{amsxtra}

\def\institute#1{\gdef\@institute{#1}}
\let\oldmaketitle\maketitle
\renewcommand\maketitle{\oldmaketitle\noindent\@institute}

\begin{document}
\def\Journal#1#2#3#4{{\em #1} #2, #3 }
\def\NCA{ Nuovo Cimento}
\def\PU{ Phys. Usp.}
\def\RNC{ Rivista Nuovo Cimento}
\def\NIM{ Nucl. Instrum. Methods}
\def\NIMA{{\em Nucl. Instrum. Methods} A}
\def\NPB{{ Nucl. Phys.} B}
\def\PLB{{ Phys. Lett.}  B}
\def\PRL{ Phys. Rev. Lett.}
\def\PRD{{ Phys. Rev.} D}
\def\ZPC{{ Z. Phys.} C}
\def\GaC{ Gravitation and Cosmology}
\def\GaCS{{ Gravitation and Cosmology} Supplement}
\def\JETP{ JETP}
\def\JETPL{ JETP Lett.}
\def\PAN{ Phys.Atom.Nucl.}
\def\CQG{ Class. Quantum Grav.}
\def\APJ{ Astrophys. J.}
\def\SCI{ Science}
\def\MPLA{{ Mod. Phys. Lett.}  A}
\def\IJTP{ Int. J. Theor. Phys.}
\def\IJMPA{{ Int. J. Mod. Phys.}  A}
\def\IJMPD{{ Int. J. Mod. Phys.}  D}
\def\NJP{ New J. of Phys.}
\def\PREV{ Phys. Rev.}
\def\PREP{ Phys. Rep.}
\def\ARAA{ Ann. Rev. Astron. Astrophys.}
\def\AIPCP{ AIP Conf. Proc.}
\def\JHEP{ JHEP}
\def\JCAP{ JCAP}
\def\EPHJ{ Eur. Phys. J.}
\def\JPCS{{ J. Phys.:} Conf. Ser.}
\def\BWP{ Bled Workshops in Physics}
\def\s{{\,\rm s}}
\def\g{{\,\rm g}}
\def\eV{\,{\rm eV}}
\def\keV{\,{\rm keV}}
\def\MeV{\,{\rm MeV}}
\def\GeV{\,{\rm GeV}}
\def\TeV{\,{\rm TeV}}
\def\sv{\left<\sigma v\right>}
\def\({\left(}
\def\){\right)}
\def\cm{{\,\rm cm}}
\def\K{{\,\rm K}}
\def\kpc{{\,\rm kpc}}
\def\beq{\begin{equation}}
\def\eeq{\end{equation}}
\def\bea{\begin{eqnarray}}
\def\eea{\end{eqnarray}}

\title{Open questions of BSM Cosmology}
\author{M.Yu.~Khlopov$^{1,2,3}$ \footnote{khlopov@sfedu.ru}}
\institute{
$^1$ Virtual Institute of Astroparticle Physics, 75018 Paris, France\\
$^2$ Center for Cosmoparticle physics ``Cosmion'' and \\ National Research Nuclear University MEPhI, 115409 Moscow, Russia\\
$^3$ Research Institute of Physics, Southern Federal University,\\ Stachki 194, Rostov on Don 344090, Russia\\
}

\maketitle

\begin{abstract}
BSM physics, on which the now standard inflationary cosmology with baryosynthesis and dark matter/energy is based, inevitably leads to cosmological scenarios beyond this
standard model, involving specific model dependent choice of models and parameters of BSM physics. Such model dependent cosmological predictions may have already found confirmations in the positive results of direct dark matter searches by DAMA/NaI and DAMA/LIBRA
experiments, interpretation of the results of Gravitational Wave experiments in terms of Primordial Black Hole merging, observation of Stochastic Gravitational Wave background by Pulsar Timing Arrays, indications of early galaxy formation in the observations of James Webb Space Telescope and searches for cosmic antihelium in the AMS02 experiment. We discuss the open questions in studies of these signatures of BSM cosmology.
\end{abstract}

\section{Introduction}\label{s:intro}
The now Standard cosmology involves inflation, baryosynthesis and dark matter/energy \cite{PPNP,Lindebook,Kolbbook,Rubakovbook1,Rubakovbook2,book,newBook,4,DMRev}. These necessary elements of the modern cosmological paradigm imply physics Beyond the Standard Model (BSM) of fundamental interactions. To probe this physics its model dependent cosmological consequences can be used as cosmological messengers \cite{Bled21,Protvino,ecu}. Such messengers lead to deviations from the standard $\Lambda$CDM cosmology. Observational evidence for such messengers would remove the conspiracy of the deviations from the standard cosmology \cite{ijmpd19}. Positive hints to such deviations need more detailed analysis of the open questions on physical basis and observable features of the corresponding BSM messengers. Here we discuss the open problems of the analysis of these hints presented at the XXVII Bled Workshop "What comes beyond the Standard models?".

If model dependent messengers of BSM physics are confirmed, it would strongly reduce the set of BSM models and their parameters. Therefore, any approach to the unified description of Nature \cite{Norma,Norma2} should inevitably include together with physical basis for inflation, baryosynthesis and dark matter also BSM cosmological signatures, which find observational support. 

We consider open questions of the dark atom interpretation of the results of the direct searches of dark matter (Section \ref{da}). We discuss the Axion Like Particle (ALP) models and the footprints of their physics in stochastic gravitational wave background (SGWB) and favored by James Webb Space Telescope (JWST) early galaxy formation. We consider open problems of ALP physics and cosmology in the scenarios of creation and evolution the primordial objects of macroscopic antimatter in our Galaxy, specifying their role as possible sources of antihelium component of cosmic rays (Section \ref{pbh}). In the conclusive Section \ref{cpp} we put the signatures and significance of the discussed messengers of BSM physics and cosmology in the context of cosmoparticle physics.
\section{Open problems of dark matter physics and cosmology}
During the last few decades the mainstream of dark matter studies was stimulated by the miracle of Weakly Interacting Massive particles (WIMPs) explaining the cosmological dark matter. Indeed, the frozen out amount of particles with mass in the hundred GeV range with annihilation cross section of the order of ordinary weak interaction could naturally explain the observed dark matter by their predicted contribution into the cosmological density. Theoretical basis for WIMP studies was found in supersymmetry (SUSY), which could naturally predict neutral WIMP-like candidate as stable lightest SUSY particle. It put experimental direct WIMP searches in the correspondence with the search for SUSY particles at the LHC. However, SUSY particles were not detected at the LHC in the hundred GeV range. This lack of positive evidence for SUSY particles can indicate very high energy SUSY scale \cite{ketovSym}. Then SUSY cannot be used to solve the internal problems of the Standard model (the divergence of Higgs boson mass and the origin of the energy scale of the electroweak symmetry breaking). It implies a non-SUSY solution of these problems. It inevitably draws attention to non-WIMP candidates for dark matter, originated from non-SUSY physical basis. 
\subsection{Dark atom probe in direct dark matter searches}\label{da}
The results of the underground direct dark matter search look controversial. Positive results of DAMA/NaI and DAMA/LIBRA experiments are presented with increasingly high statistics \cite{rita}. It seems to be in sharp contradiction with negative results of other groups (see \cite{PPNP} for review and references). However, the strategy of most of these groups is aimed to the WIMP search. Though the results of DAMA/NaI and DAMA/LIBRA experiments, taken separately admit their WIMP interpretation \cite{rita}, confrontation with publications of other groups strongly favors non-WIMP interpretation of these positive results and appeals to consider them as the experimental evidence for dark atom nature of dark matter \cite{PPNP,ecu,I,kuksa}.

The dark atom hypothesis assumes existence of stable particles with negative even electric charge $-2n$. Such particles should be generated in excess over their antiparticles and bind with $n$ primordial helium-4 nuclei, created in Big Bang Nucleosynthesis, in neutral nuclear interacting atom like states. 

Multiple charged stable states can be related with the composite nature of Higgs boson. If Higgs boson constituents are charged, their composite multiple charged states can exist. By construction Higgs boson constituents possess electroweak charges. It can not only provide prediction of stable $-2n$ charged particles, but also can give rise to their excess over positively charged antiparticles owing to electroweak sphaleron transitions, which balance this excess with baryon asymmetry. Such balance takes place in Walking Technicolor model (WTC) and the excess of $-2n$ charged particles as the constituents of dark atoms determines the relationship between dark matter density in the form of dark atoms and baryon density. The only parameter of dark atom model is the mass of the stable $-2n$ charged particles. This parameter can be determined under the assumption that dark atoms explain all the observed  dark matter density   \cite{arnab,sopin,Bikbaev_2024}. 

It was shown in \cite{arnab,sopin,Bikbaev_2024} that the excess of $-2n$ charged particles balanced with baryon excess can be generated in the case of any new particle family, which possess electroweak charges. It makes possible to balance with baryon asymmetry the excess of -2 charged ($\bar U \bar U \bar U$) clusters of stable antiquarks $\bar U$ with the charge $-2/3$. Such a new stable generation is predicted as the 5th generation in the approach \cite{Norma2}. The excess generated by the sphaleron balance depends on the mass of multiple charged particles. It can explain the observed dark matter, if the mass of stable multiple charged particles doesn't exceed few TeV. It puts upper limit on the mass of multiple charged particles, at which dark atoms can explain the observed density of dark matter,  and challenges their experimental search at the LHC \cite{sopin}.

The problems of dark atom structure and interaction are more close to the nuclear, than to atomic physics. Instead of small nuclear interacting core covered by large leptonic (electronic) shell, multiple charged lepton-like core of dark atom is closely covered by nuclear interacting helium shell of nuclear size.  

The structure of dark atom and its properties tend to a nuclearite (Thomson like atom) -- to a specific superheavy neutral nuclear matter species. 

Bohr-like $O$He atom description can be appropriate only for a double charged particle ($n=1$) bound with primordial helium nucleus. Even in this case radius of the helium Bohr orbit is nearly equal to the size of this nucleus. The $-2n$ charge particle forms at $n>1$ Thomson-like $X$He atom with $n$ helium nuclei. $-2n$ charged lepton is situated in this atom within an $n$-$\alpha$-particle nucleus. It makes $X$He dark atoms more close to $O$-nuclearites \cite{gani}, in which electric charge of nuclear matter is compensated by negative charge of heavy leptons. 

Evolution of these species formed in the period of Big Bang Nucleosynthesis is the open problem of dark atom scenario. $X$He atoms are formed in the medium enriched by free helium nuclei and their binding with $X$He can lead to production of anomalous isotopes, which is now the object of our thorough investigation. The role of the Bose-Einstein nature of $\alpha$ particles is of particular interest in this investigation.

Negative results of direct WIMP searches find natural qualitative explanation by the Dark atom hypothesis. Dark atom nuclear interaction in the terrestrial matter causes their slowing down, which results in negligible nuclear recoil in the dark atom collisions within underground detectors \cite{PPNP,ecu,I}. 

Local dark atom concentration is determined at each level of terrestrial matter by the balance between the incoming cosmic flux of dark atoms and their diffusion towards the center of Earth. Such concentration is adjusted to the incoming cosmic flux. At the 1 km depth the equilibrium is maintained at the timescale of less than 1 hour. Since the incoming flux possess annual modulations due to the Earth's orbital motion around the Sun, the dark atom concentration experiences annual modulation within the matter of the underground detector.  If dark atoms can form low energy (few keV) bound states with nuclei of detector, the energy release in such binding should possess annual modulation. It can explain the signal, detected in DAMA/NaI and DAMA/LIBRA experiments, assuming that the dark atom number density in the detector is adjusted to the incoming flux, that there is a 3 keV bound state of dark atom with sodium nucleus and that the rate of radiative capture to this bound state is determined by the E1 transition with the account for isospin symmetry breaking factor (given by the ratio of the neutron and proton mass difference to the nucleon mass). Under these assumptions the results of DAMA/NaI and DAMA/LIBRA can be explained and it challenges the analysis of these experimental data \cite{rita} in the framework of the dark atom hypothesis.

The basic open problem of the dark atom hypothesis is related to the proper quantum-mechanical description of the nuclear interaction of dark atoms. In the lack of small parameters, used in the ordinary atomic physics, the numerical methods of continuous approach to the description of three body problem of dark atom constituents+nucleus were developed \cite{timur}.  These methods were elaborated for interaction of nuclei with both Bohr-like and Thomson-like dark atoms  \cite{timur2,timur3} (see \cite{Bikbaev_2024} for recent review and references). Development of the proper quantum mechanical description of dark atom interaction with nuclei is necessary for analysis of dark atom formation of anomalous isotopes after BBN, of effects of dark atom capture by stars and effects in stellar nucleosynthesis, as well as for the proof of dark atom interpretation of DAMA/NaI and DAMA/LIBRA results.  
\subsection{Dark Matter in Axion-Like Particle Models}
Axion-Like Particle (ALP) models go far beyond the original axion models and their relationship with the Peccei-Quinn solution for the problem of strong CP violation in QCD. ALP can be reduced to a simple model of a complex pseudo-Nambu-Goldstone field \beq \Psi = \psi \exp{i \theta} \label{psi}\eeq with broken global $U(1)$ symmetry \cite{PPNP,Protvino}. The potential
$$V=V_0+\delta V$$
 leads to spontaneous breaking of the $U(1)$ symmetry by the term 
 \beq V_0= \frac{\lambda}{2}(\Psi^{\ast} \Psi -f^2)^2, \label{vo} \eeq
which retains continuous (phase) degeneracy of the ground state
\beq \Psi_{vac}= f \exp (i \theta) \label{cav} \eeq
 and  to manifest breaking of the residual symmetry by the term 
 \beq \delta V (\theta)= \Lambda^4 (1-\cos\theta) \label{disc} \eeq\
 with $\Lambda \ll f$, leading to a discrete set of degenerated ground states, corresponding to 
$$\theta_{vac} = 0, 2\pi, 4\pi, ...$$ 
In the result of the symmetry breaking by the term (\ref{disc}) an ALP field $\phi = f \theta$ is generated with the mass \beq m_{\phi} = \Lambda^2/f. \label{pngm} \eeq
 
 In the axion models the term (\ref{disc}) is generated by instanton transitions. In ALP models, it is present in the theory initially. When Hubble parameter decreases down to the value of the ALP mass, given by the Eq. (\ref{pngm}), $H= m_{\phi} = \Lambda^2/f$,  coherent oscillations of the ALP field start,  The ALP field energy density is proportional to $\theta_i^2$, where $\theta_i$ is the amplitude of ALP field oscillations, given by the initial local value of the phase, when the oscillation starts. 
 
 The ALP field represents the Bose-Einstein condensate of ALP Bose-gas in its ground state. Therefore. in spite of a very small mass, ALP in the condensate are nonrelativistic and, if they dominate at the matter dominated stage, can play the role Cold Dark matter, reproducing the Standard $\Lambda$CDM scenario of Large Scale Structure formation. However, ALP physics can lead to strong deviations of the Standard cosmological paradigm and to creation of strong primordial inhomogeneities of different kinds, discussed in the next section \ref{pbh}, .

\section{BSM cosmology from ALP physics}\label{pbh}
If the first phase transition takes place after reheating, the correlation radius is small and ALP strings are formed. The string network is converted into unstable walls-surrounded by strings structure after the second phase transition. ALP energy density distribution represents the replica of this structure and preserves large scale correlations in the nonhomogeneity of the ALP energy density \cite{PPNP}. Evolution of this primordial inhomogeneity, its role in structure formation and observational signatures of the corresponding scenario are the open questions of this direction of the ALP studies. The question of the relationship of this large scale structure and formation of axion stars \cite{tkacheva} is of special interest. 

After the first phase transition, if it takes place at the inflationary stage, the phase has a fixed value within the Hubble radius at each e-folding. Quantum fluctuations lead to variations in phase between disconnected regions. These fluctuations are given by \beq \delta\theta = \frac{H_{\rm infl}}{2\pi f},\label{ph}\eeq where $H_{\rm infl}$ is Hubble parameter at the inflationary stage \cite{PPNP}. After the second phase transition the local value of phase determines the amplitude $\theta - \theta_{vac}$ of the ALP field oscillations. If at the e-folding, corresponding to the observed part of the Universe, $\theta <\pi$ 
$\theta_{vac} = 0$. In regions, where the fluctuations (\ref{ph}) moved the phase at successive steps of inflation to values greater than $\pi$,  $\theta_{vac} = 2\pi$. At the border of domains  with  $\theta_{vac} = 2\pi$ and surrounding regions with $\theta_{vac} = 0$ along the closed surfaces with $\theta = \pi$ closed domain walls are formed in the course of the second phase transition. 

\begin{figure}
    \begin{center}
        \includegraphics[scale=0.6]{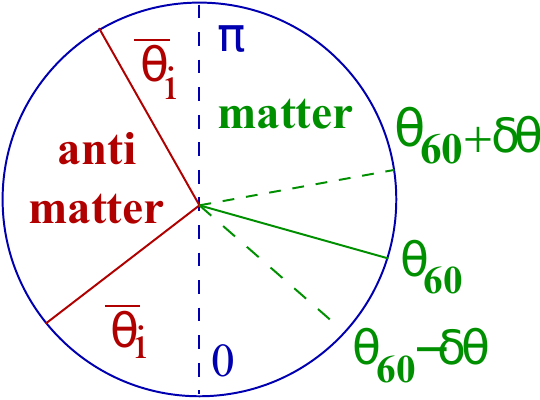}
        \caption{Phase fluctuations at inflationary stage can cross $\pi$. It leads to formation of closed domain walls. If the ALP field is unstable relative to decays to quarks and leptons with baryon and lepton number nonconservation, as it is the case in the model of spontaneous baryosynthesis \cite{DolgovAM}, crossing $\pi$ and 0, results in formation of antibaryon domains in baryon asymmetric Universe \cite{PPNP} }
        \label{a}
    \end{center}
\end{figure}

If the field $\Psi$ interacts with quarks and leptons with nonconservation of the baryon and lepton numbers, decay of the field $f\theta$ generate baryon asymmetry in its motion to the ground state. If $\theta < \pi$, baryon excess is generated, while the antibaryon excess is generated in domains with $\theta > \pi$ (see Fig. \ref{a}).

The open question in the scenario with the first phase transition at the inflationary stage is the necessity to suppress large scale fluctuations, excluded by the observed  CMB isotropy. One of the possible solution can be a large amplitude $\phi$ of the complex ALP field (\ref{psi}) at the stage of inflation, corresponding to the modern horizon, and its decrease down to the broken symmetry value $f$ (see Eq.(\ref{cav})) at successive inflationary stages.
\subsection{PBH, SGWB and JWST signatures of ALP physics}
Inflation provides homogeneous and isotropic initial conditions for evolution of causally disconnected regions within the observed cosmological horizon. However, such evolution can lead to qualitative difference of conditions in different regions. 
Formation of closed domain walls, related with the difference in the ground states of the ALP field gives several examples of such kind. Pending on the two fundamental scales of the ALP physics, $f$ and $\Lambda$ it can lead to three possibilities.

The interval of mass $M$  of domain walls, collapsing in primordial black holes (PBH) \cite{ketovSym,PBHrev}, is given by 
\beq M_{min} = f\left(\frac{m_{Pl}}{\Lambda} \right)^2 \le M \le M_{max} = f \left(\frac{m_{Pl}}{f}\right)^2 \left(\frac{m_{Pl}}{\Lambda}\right)^2. \label{pbhalp}\eeq
Here the minimal mass $M_{min}$ is determined by the condition that the gravitational radius of wall $r_g = 2 M/m_{Pl}^2$ exceeds the width of wall $d \sim f/\Lambda^2$. The maximal mass corresponds to the condition that the wall as a whole enters horizon before it starts to dominate within it. Small walls with mass $M< M_{min}$ form oscillons, while large walls with mass $M>M_{max}$ separate the region of their dominance from the other part of the Universe, 

At the appropriate value of $f$ and $\Lambda$ the ALP mechanisms can provide formation of PBHs with stellar mass, and even larger, up to the values corresponding to the seeds for Super Massive Black Holes in Active Galactic nuclei (AGN) \cite{AGN,RubinCluster,dolgovPBH}. Their mass can easily exceed the limit of pair instability and the LIGO/VIRGO detected gravitational wave signal from coalescence of black holes with masses $M > 50 M_{\odot}$  naturally puts forward the question on their primordial origin \cite{LV150,LV150apl}. 

One can consider recent discovery of Stochastic Gravitational Wave Background (SGWB) by Pulsar Timing Arrays (PTA) \cite{nano} as another evidence of ALP physics. 

Large closed domain walls with mass $M>M_{max}$ separate the region of their dominance from the surrounding Universe. Formation of these walls is accompanied by gravitational wave background radiation, which can reproduce the PTA data \cite{SER}. In the ALP model, the contour of the future domain wall is created, when the phase crosses $\pi$. The values of phase at the stages of inflation, preceding the this crossing,  approach to $\pi$. In the result the ALP energy density in the regions surrounding future domain wall is much larger, than the average one for the ALP field. Therefore, when the wall separates from the other part of the Universe the region, where it dominates, the ALP density in the surrounding regions is much higher than the average in the Universe and the galaxy formation in these regions can take place much earlier, than in the rest of the Universe and can happen  at the redshifts $z>10$, as indicated by the data of JWST \cite{JWST}. In that way ALP physics can simultaneously explain the PTA and JWST data \cite{shu}. The open questions in this scenario are generation of GW background from the domain walls and evolution of the regions of the enhanced ALP density. In particular, the question is of special interest whether black hole formation is possible in their central part, or the enhanced ALP density itself plays the role of AGN seeds in the early galaxies. 

Evolution of wall dominated regions is also of special interest. Their appearance in the Lemaitre-Friedman-Robertson-Walker Universe, supported by inflational scenario, is the effect of fluctuations at the inflationary stage, resulted in the strong deviation of some regions from isotropy and homogeneity. The influence of such regions on the evolution of surrounding regions needs special study.

\subsection{Antimatter domains in baryon asymmetric Universe}
Baryon asymmetry of the Universe reflects the absence in the observed Universe of macroscopic antimatter in the amount comparable with the amount of baryonic matter. In the now standard cosmological paradigm baryon asymmetry is related to baryosynthesis, in which baryon excess is created in very early Universe. If  baryosynthesis is inhomogeneous, in the extreme case the sign of the baryon excess can change, giving rise to antimatter domains, produced in the same process, in which the baryon asymmetry is created \cite{CKSZ,DolgovAM,Dolgov2,Dolgov3,KRS2,AMS}. Such antimatter domains are surrounded by matter. They should be sufficiently large to survive to the present time and to give rise to antimatter objects in the Galaxy. It implies also effect of inflation in addition to nonhomogeneous baryosynthesis. Such combination of inflation and nonhomogeneous baryosynthesis can take place in the ALP model of spontaneous baryosynthesis  \cite{DolgovAM} at the specific choice of its parameters. 

This choice determines the properties of antimatter domains, their evolution and the forms of celestial antimatter objects in baryon asymmetrical Universe.  The minimal mass of the surviving domain is determined by its annihilation with the surrounding matter. The upper limit on the possible amount of antimatter in our Galaxy follows from the observed gamma background \cite{AMS,GC}. These limits give the interval of mass $M$ of antimatter in our Galaxy  \beq 10^3 M_{\odot} \le M \le 10^5 M_{\odot},\label{gc}\eeq which is typical for globular clusters.  Symmetry of electromagnetic and nuclear interactions of matter and antimatter would make celestial antimatter objects looking like matter ones. Globular clusters are situated in the halo of our Galaxy, where matter gas density is low. It seemed to favor the hypothesis of antimatter globular cluster in our Galaxy, which may be rather faint gamma source. If antibaryon density is much higher, than the baryonic density, specific ultra-dense antibaryon stars can be formed \cite{Blinnikov}, 

Therefore, studies of possible forms of macroscopic antimatter objects in our Galaxy are challenging, involving evolution of antibaryon domains in baryon asymmetrical universe \cite{orch,orch2} in the context of models of nonhomogeneous baryosynthesis. 

The idea of antimatter globular cluster inspired to look for an appropriate galactic Globular cluster as its possible prototype. The observed properties of the M4 globular cluster were used for this purpose \cite{nastya}, but such approach implies strong correction. Indeed, chemical evolution within the isolated antimatter domain cannot be the same as that for the ordinary matter. At rather wide range of antibaryon density one can expect that primordial nucleosynthesis in the domain should lead to production of primordial antihelium. However, products of anti-stellar nucleosynthesis can neither come to the domain from other parts of the Galaxy, nor remain in domain being produced by antimatter stars within it. It makes highly improbable enrichment of antimatter object by metallicity, while all the observed galactic globular clusters don't have metallicity below the Solar one. It demonstrates strong mixture of products of stellar nucleosynthesis over the Galaxy, which is not possible for the chemical evolution of antimatter within the isolated region of antibaryon domain.

Primordial metallicity can be produced in domains with high antibaryon density. In the context of spontaneous baryosynthesis based on ALP physics such high density antibaryon domains can correspond to crossing $\pi$. It should be accompanied by massive domain walls at their border. Such walls with the mass $$M > M_{max} = f \left(\frac{m_{Pl}}{f}\right)^2 \left(\frac{m_{Pl}}{\Lambda}\right)^2 = \left(\frac{m_{Pl}}{f}\right)^2 M_{min}$$ would dominate in the corresponding region, separating it from the rest of the Universe. In such scenario only domains surrounded by walls with $M < M_{max}$ can leave the antibaryon domain observable. It puts an open question, whether such domains are sufficiently large to survive in the matter surrounding and what is the result of the domain wall evolution.  

The  sensitivity of AMS02 experiment is far below the predicted flux of cosmic antihelium from astrophysical sources  \cite{poulin}. Since 2017 a  suspected antihelium-4 event is demonstrated at seminars, but remains unpublished. The unprecedented sensitivity of AMS02 experiment makes the collaboration especially responsible for publication of its results. That is why all the possible background interpretation of such candidates should be checked  before the discovery of cosmic antihelium-4 is announced.

In any case to confront AMS02 searches the prediction for composition and spectrum of cosmic antinuclei from antimatter objects in our Galaxy should be made. Such prediction inevitably involves the account for propagation of antinuclei in galactic magnetic fields \cite{nastya2} as well as for the inelastic processes in such propagation. 
\section{Conclusions}\label{cpp}
The results of DAMA experiments, LIGO-VIRGO-KAGRA, PTA and JWST data, possible existence of antihelium component of cosmic rays increase the hints to new physics phenomena. Their confirmation will manifest deviations from the standard cosmological model. BSM cosmology, involving Warmer-than-Cold dark matter scenario of nuclear interacting dark atoms of dark matter, or primordial strong nonhomogeneities of energy and/or baryon density, can give rise to new scenarios of galaxy formation and evolution.  We have outlined here the open questions in the proposed BSM models and scenarios. Answers to these questions need special studies and will deserve discussion at future Bled Workshops.

In the context of cosmoparticle physics, confirmation of the existence of cosmological messengers of new physics would provide a sensitive probe for BSM cosmology and for specific choice of BSM models and their parameters, on which the BSM scenario is based. In this case only such models, which predict the detected deviations from the standard cosmological paradigm can pretend to be realistic. The hints to the cosmological messengers of BSM physics may appeal to reanalysis of the observational constraints and interpretation of the edges of such constraints in the terms of the BSM cosmology. The observed accelerated expansion of the modern Universe may challenge our paradigm of UNIverse, putting us into a baby universe separated from the mother multiverse and appealing to development of special methods to study such cosmology.
\section*{Acknowledgements}
The research was carried out in Southern Federal University with financial support of the Ministry of Science and Higher Education of the Russian Federation (State contract GZ0110/23-10-IF).



\end{document}